\def\cm{cm$^{-1}$}
\documentclass[final,5p,times,twocolumn]{elsarticle}
\date{March 18, 2010}
\usepackage{graphicx}
\usepackage{amssymb}

\journal{Journal of Physics and Chemistry of Solids}

\begin{document}

\begin{frontmatter}

\title{Looking at the superconducting gap of iron pnictides}


\author[1]{Martin Dressel}\corref{cor1}\ead{dressel@pi1.physik.uni-stuttgart.de}
\author[1]{Dan Wu}
\author[1]{Neven Bari\v{s}i\'{c}}
\author[1,2]{Boris Gorshunov}

\address[1]{1. Physikalisches Institut, Universit\"at Stuttgart, Pfaffenwaldring 57, 70550 Stuttgart, Germany}
\address[2]{Prokhorov Institute of General Physics, Russian Academy of Sciences,  Vavilov str. 38, 119991 Moscow, Russia}

\cortext[cor1]{Corresponding author.}

\begin{abstract}
THz and infrared spectroscopy is widely utilized to investigate
the electrodynamic properties of the novel iron-based superconductors in the normal and superconducting states. Besides
electronic excitations and correlations, electron-phonon coupling
and the influence of magnetism, the experiments yield important
information on low-lying excitations and help to clarify the
number and symmetry of superconducting gaps. While the
experimental data of different groups converge, the interpretation
is still under debate. Here we review the status of optical
investigations on the superconducting state for the 122 and 11
family of iron pnictides.
\end{abstract}

\begin{keyword}
Superconducting energy gap
\sep
Iron pnictides
\sep
Optical spectroscopy
\sep
Order parameter

\PACS
74.25.Gz    %Optical properties of superconductors
\sep
74.70.Xa  %Pnictides and chalcogenides
\sep
78.20.-e    %Optical properties of bulk materials and thin films

\end{keyword}

\end{frontmatter}



\section{Introduction}
Macroscopically superconductivity is characterized by vanishing
resistivity and perfect diamagnetism; microscopically, a gap in
the density of states opens at the Fermi surface \cite{Tinkham96}.
Both phenomena are subject of the BCS theory which assigns an
order parameter to the superconducting state. This immediately
gives some insight into the symmetry and eventually the mechanism
of superconductivity \cite{Schrieffer99}. It works almost
perfectly in conventional single band metallic superconductors, it has some
success in the description of more exotic systems
\cite{Ketterson08}, such as heavy fermions \cite{Ott87} or organic
superconductors \cite{Ishiguro98}, but requires care when applied to high-temperature superconductivity in transition metal
oxides.

When the new class of iron pnictides was discovered two years ago
\cite{Kamihara08}, immediately the questions of mechanism and
symmetry arose. Despite the enormous efforts undertaken of the
entire community \cite{Ishida09} the present answers are not
completely satisfactory.

Since the development of the BCS theory,  optical investigations
contributed a great deal to solve the puzzle of
superconductivity. And for each new family of exotic compounds
discovered over the decades, high-frequency studies and infrared
spectroscopy were one of the first and decisive investigations
frequently pointing the direction for further studies of
complementary methods \cite{Dressel04,Basov05,Dressel08,Calvani08}.

Following these lines, optical methods were quickly applied to iron pnictides and revealed the spin-density-wave gap in parent compounds, as well as the
electronic properties of the superconducting systems \cite{Chen08,Chen09,Dong08,Hu08,Li08SC,Pfuner09,Hu09c,Yang09,Wu09,Nakajima09,Qazilbash09,Hu09n,Akrap09}.

By now there is only a limited number of optical investigations on
the superconducting properties of iron-pnictides. Early studies on
non-crystalline materials gave some rough ideas
\cite{Chen08,Chen09l,Dubroka08,Drechsler08,Mirzaei08}, but could
not contribute to the main questions posed above. Only very
recently, several groups published their findings on single
crystals of the 122 family
\cite{Li08SC,Wu10b,Wu10c,Kim09,Heumen09} and 11 compound
\cite{Homes10}. Also epitaxial grown thin films have been explored
\cite{Gorshunov10,Perucchi10}. As we will review in this
contribution, the optical data look very similar, but
interestingly, the interpretation is very dispersed.

\section{Results and Discussion}
\subsection{Hole-doped 122 iron pnictides}
\begin{figure}
 \includegraphics[width=0.8\columnwidth]{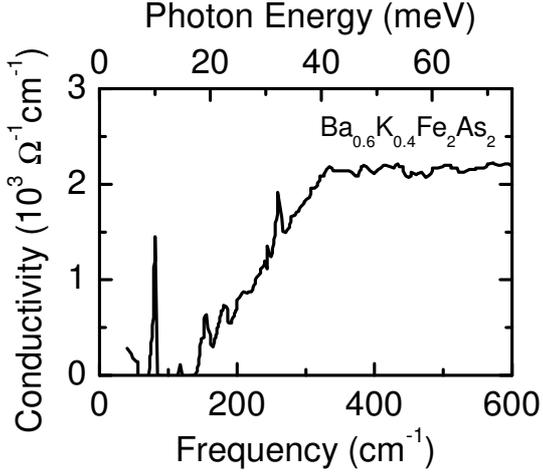}
 \caption{Frequency dependent conductivity of Ba$_{0.6}$K$_{0.4}$Fe$_2$As$_2$  obtained from reflectivity measurements at $T=10$~K. The data are taken from Li {\it et al.} \cite{Li08SC}.
\label{fig:hole}}
 \end{figure}

In very early experiments on the hole-doped
compound Ba$_{0.6}$K$_{0.4}$Fe$_2$As$_2$, Li {\it et al.}\cite{Li08SC} found clear indications of the
opening of a complete gap. The reflectivity shows a sharp upturn
below $T_c$ which approaches unity around 150~\cm; the optical
conductivity drops to zero in a linear fashion as demonstrated in
Fig.~\ref{fig:hole}. The behavior can be fitted by assuming a
single $s$-wave gap with $2\Delta_0/h c = 110$~\cm.\footnote{Here
$h= 2\pi \hbar=  4.1 \times 10^{-15}$~eV\,s is the Planck constant, and
$c=3\times 10^8$~m/s is the speed of light. Note that optical
experiments probe excitations across the full gap of $2\Delta$.} A
fit by two gaps ($2\Delta_0^{(2)}/hc=110$~\cm\ and
$2\Delta_0^{(1)}/hc=190$~\cm) gives a slightly better description
\cite{Hu09c}. From the spectral weight removed below the gap, the
superconducting penetration depth of
Ba$_{0.6}$K$_{0.4}$Fe$_2$As$_2$  can be calculated:
$\lambda=2000$~\AA. The spectral weight analysis shows that the
Ferrell-Glover-Tinkham sum rule \cite{Tinkham96,DresselGruner02}
is satisfied below $6\Delta_0$.

\subsection{Electron-doped 122 iron pnictides}
\subsubsection{Films}
Comprehensive optical investigation of
Ba\-(Fe$_{0.9}$\-Co$_{0.1})_2$As$_{2}$ thin films  in a wide
frequency and temperature range elucidate the electrodynamic
properties in the superconducting state. A 900~\AA\ thick film with $T_c=20$~K was laser-deposited on a (La,Sr)(Al,Ta)O$_3$ substrate.
Using a Mach-Zehnder interferometer Gorshunov {\it et al.}
\cite{Gorshunov10} were able to directly observe the opening of
the superconducting gap of $2\Delta_0=3.7$~meV corresponding to 30~\cm, i.e.\
$2\Delta_0/k_BT_c=2.1 \pm 10\%$ by THz transmission and phase shift
measurements. In this region
the temperature and frequency dependence of the conductivity is
well described by the BCS theory assuming a complete isotropic
gap, as plotted in Fig.~\ref{fig:electron}(a). However, there
remains a strong quasi-particle absorption below 10~\cm, which is
not predicted by the simple model. The spectral weight of the
condensate $1.94\times 10^7~{\rm cm}^{-2}$ corresponds to a
penetration depth $\lambda=3600$~\AA.

Using synchrotron radiation, Perucchi {\it et al.} measured the
THz reflectivity of a high quality epitaxial thin film of
Ba\-(Fe$_{0.92}$\-Co$_{0.08})_2$As$_{2}$ (thickness $d=3500$~\AA\
with $T_c=22.5$~K) on a DyScO$_3$ substrate with an epitaxial
SrTiO$_3$ intermediate layer \cite{Perucchi10}. They did not
evaluate the intrinsic material properties, but only the composite
reflectance of film and substrate, and presented the ratio of
superconducting and normal states: $R_s(T)/R_n$. For that reason
no conductivity could be calculated and plotted here. Nevertheless,
clear evidence for a superconducting gap $2\Delta_0^{(1)}/hc
\approx (30\pm 1)$~\cm\ was observed. Following the approach of
other groups and methods, a fit by a two-band, two-gap model
yields better agreement. The larger gap is estimated to be at
approximately [$2\Delta_0^{(2)}/hc=(110\pm 15)$~\cm]. The results
are seen as evidence for a nodeless isotropic double-gap scenario,
with the presence of two optical gaps corresponding to
$2\Delta_0/k_BT_c$ values close to 2 and 7. The two contributions
to the optical conductivity have a plasma frequency of $10^4$~\cm\
and scattering rates of 300 and 1200~\cm, respectively.

\subsubsection{Single Crystals}
\begin{figure}
 \includegraphics[width=0.8\columnwidth]{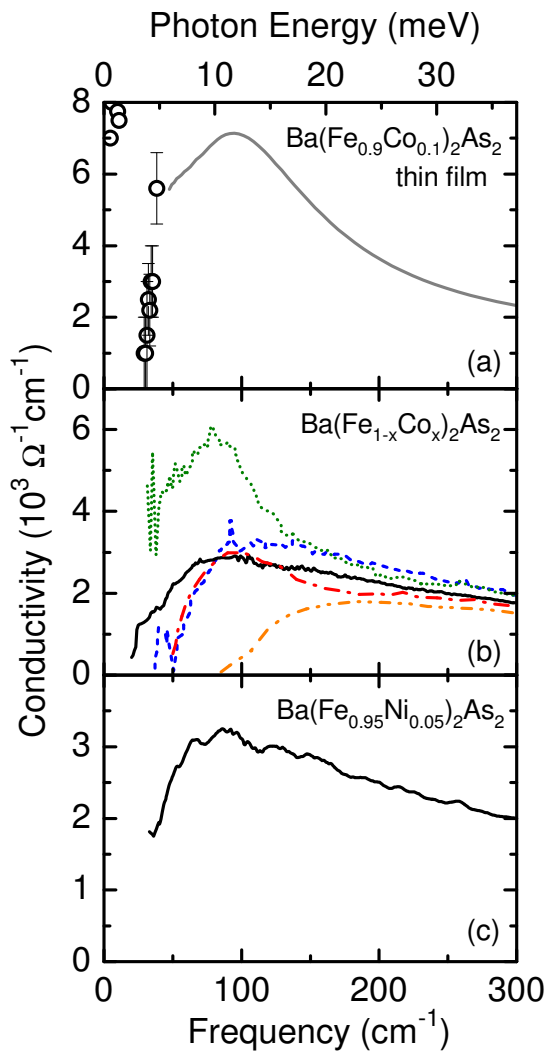}
 \caption{Optical conductivity of electron doped
 Ba\-(Fe$_{1-x}$\-$M$$_x$)$_2$As$_{2}$ close to optimal doping; $M$=Co, Ni. \label{fig:electron}
(a)~From THz transmission and phase measurements through  a thin
film of Ba\-(Fe$_{0.9}$\-Co$_{0.1})_2$As$_{2}$ at $T=5$~K (open dots) the
complete opening of the gap in conductivity can be seen around
30~\cm; the grey line is obtained by reflectivity measurements.
(b)~Reflectivity measurements performed by different
groups yield a very similar behavior at low temperatures. The
solid line was obtained by Wu {\it et al.} \cite{Wu10b,Wu10c} on
Ba\-(Fe$_{0.92}$\-Co$_{0.08})_2$As$_{2}$ at $T=9$~K. Kim {\it et
al.} \cite{Kim09} investigated
Ba\-(Fe$_{0.935}$\-Co$_{0.065})_2$As$_{2}$ at $T=5$~K (blue dashed
curve). The green dotted curve corresponds to a measurement of van
Heumen {\it et al.} \cite{Heumen09} on
Ba\-(Fe$_{0.93}$\-Co$_{0.07})_2$As$_{2}$ at $T=10$~K. 
Nakajima {\it et al.} \cite{Nakajima09} investigated 
Ba\-(Fe$_{0.94}$\-Co$_{0.06})_2$As$_{2}$ (dot-dashed  red line) and Ba\-(Fe$_{0.92}$\-Co$_{0.08})_2$As$_{2}$ (dot-dot-dashed orange line) single crystals with $T_c=25$ and 20~K, respectively. (c)~The
behavior of the Ba\-(Fe$_{0.95}$\-Ni$_{0.05})_2$As$_{2}$ at
$T=10$~K is very similar to the Co doped sister compound
\cite{Wu10b}.}
 \end{figure}

Reflectivity measurements on single crystals suffer from small
crystal  size and limited surface quality. Nevertheless it is
quite remarkable how well the data of different groups coincide,
as demonstrated in Fig.~\ref{fig:electron}(b) which shows the
optical conductivity as obtained by Kramers-Kronig analysis as a
function of frequency for the lowest temperature available. In all
cases significant deviations from the normal state behavior occur
below 200~\cm, which corresponds to approximately $6-8\Delta_0$. In the superconducting phase the conductivity
passes through a maximum around 80~\cm, before it drops.
There seems to be a considerable in-gap absorption present
as a more gradual reduction with decreasing frequency
than expected for a simple $s$-wave superconductor.

The first broadband investigations on single crystals of
electron-doped iron pnictides were performed by Wu {\it et al.}
\cite{Wu10b,Wu10c}. They measured
Ba\-(Fe$_{0.92}$\-Co$_{0.08})_2$As$_{2}$ in a wide temperature
range and frequencies from 20~\cm\ to 30\,000~\cm. Upon
passing the superconducting transition at $T_c=25$~K, the
reflectivity rises toward unity, leading to a clear gap-like
feature in the conductivity as depicted in
Fig.~\ref{fig:electron}(b)  as a solid line. The conductivity
below 100~\cm\ drops due to the complete opening a gap in the
density of states at $2\Delta_0^{(1)}/hc=50$~\cm, corresponding to
$2\Delta_0/k_BT_c\approx 2.5-3$. It can be well described by the
BCS theory with no nodes assumed in the order parameter. There
remains a considerable contribution below which infers that not
all carriers are gapped. Alternatively, a superior fit is
suggested with a second gap at 17~\cm, following suggestions by
APRES that two gaps are present
\cite{Evtushinsky09,Khasanov09,Mazin08}. The missing spectral
weight amounts to a penetration depth of $\lambda=(3500\pm 350)$~\AA\ for
Ba\-(Fe$_{0.92}$\-Co$_{0.08})_2$As$_{2}$.

The behavior of Ba\-(Fe$_{0.95}$\-Ni$_{0.05})_2$As$_{2}$ is very
similar to the optimally doped Co samples \cite{Wu10b,Wu10c}. The
reflectivity turns up around 60~\cm\ and approaches unity in way
that the optical conductivity becomes gapped with
$2\Delta_0^{(1)}/hc=35$~\cm. In Fig.~\ref{fig:electron}(c) the frequency
dependent conductivity is plotted for $T=10$~K, i.e.\ well into
the superconducting state ($T_c=20$~K). The larger carrier density
leads to a slightly smaller penetration depth of $\lambda =
(3000\pm 300)$~\AA.

Nakajima {\it et al.} presented a comprehensive study of several
Ba\-(Fe$_{1-x}$\-Co$_{x})_2$As$_{2}$ compounds with $x=0$, 0.04, 0.06, and 0.08, where the 
latter two become superconducting at $T_c=25$ and 20~K \cite{Nakajima09}. The normal state conductivity is very well described by two Drude components with very different scattering rates, as suggested by Wu {\it et al.} \cite{Wu10b}. The narrow Drude term becomes stronger in spectral weight with increasing the Co concentration $x$; the broad incoherent background is assigned to
the segments of the Fermi surface which fulfill the nesting conditions. From the rise in 
reflectivity and the corresponding suppression of $\sigma_1(\omega)$ at the lowest temperature ($T=5$~K), they conclude a full gap superconductor of $s$-wave symmetry with $2\Delta/h c \approx 80(50)$~\cm, which yields $2\Delta/k_BT_c=4.6$. No conclusion can be drawn about a second gap above or below this major feature.
As pointed out, the gap seems to develop in every piece of the Fermi surface in the superconducting state since the narrow and the broad Drude term are equally affected by the superconducting gap.
From the missing area, the penetration depth $\lambda=(2770\pm250)$ and $(3150\pm300)$~\AA\ is estimated, for the two compounds. 

Temperature dependent reflectivity measurements on
Ba\-(Fe$_{1-x}$\-Co$_{x})_2$As$_{2}$ were also performed by van
Heumen {\it et al.} \cite{Heumen09}. In Fig.~\ref{fig:electron}(b)
we present the data obtained on
Ba\-(Fe$_{0.93}$\-Co$_{0.07})_2$As$_{2}$ at $T=10$~K as dotted line. In contrast
to other groups, they find a strong peak around 80~\cm\ in the
conductivity spectrum which is assigned to interband transitions;
this makes the separation of the free charge carrier conductivity
difficult. In the course of their analysis of the missing area the
superfluid density was estimated to $(2.2 \pm 0.5)\times
10^7$~cm$^{-2}$. Plotting $-\omega^2\epsilon_1(\omega)$ shows a
minimum around 50~\cm, which indicates a gap of $2\Delta_0^{(1)}\approx
6.2$~meV with $s$-wave symmetry \cite{Marsiglio96}.
Although no other gap structure is apparent in the data,
van Heumen {\it et al.} decomposed the conductivity in a
approach similar to \cite{Wu10b} and found them to be
consistent with a second gap around $2\Delta_0^{(2)}=14$~meV as suggest by
ARPES and STS experiments \cite{Terashima09,Massee09}.

The optical properties of Co-doped BaFe$_2$As$_2$ were also measured by Kim {\it et al.}  \cite{Kim09} at temperatures  $T \geq 5$~K. The conductivity of a single crystal of Ba\-(Fe$_{0.935}$\-Co$_{0.065})_2$As$_{2}$ ($T_c=24.5$~K) as derived by Kramers-Kronig analysis of reflectivity data  down to 35~\cm\ is plotted in Fig.~\ref{fig:electron}(b) as dashed line.
In the normal state two Drude contributions [with $1/(2\pi c\tau)=90$ and 300~\cm, respectively] and some mid-infrared Lorentzians are sufficient to describe the frequency-dependent conductivity; in agreement to the approach the other groups have chosen. Also in the superconducting state
their findings are in accord with previous experiments by Wu {\it et al.} \cite{Wu10b} (solid line) as far as the overall behavior is concerned.
However, Kim {\it et al.} point out that a multi-gap scenario with at least three energy gaps ($2\Delta/k_BT_c=3.1$, 4.7 and ~9.2) is necessary to reproduce the experimental data. While the drop of $\sigma(\omega)$ to zero determines the gap $2\Delta_0^{(1)}/hc=53$~\cm\ unambiguously, it is hard to identify  gaps at 80 and 157~\cm\ in their raw data. There remains a strong in-gap absorption which becomes more pronounced at $T=20$~K and by far exceeds the predictions of a complete $s$-wave scenario.
While the gap $\Delta^{(1)}$ opens in the narrow Drude, the 80~\cm\ gap is assigned to
the broad Drude contribution.
It is interesting to note the common trend, that the larger the gap is, the smaller the spectral weight that contributes to the corresponding conductivity. The
penetration depth $\lambda=2700$~\AA\ is estimated in their report.

In summary, as already pointed out in Ref.~\cite{Perucchi10}, most
groups measuring reflectivity of electron-doped single crystals agree on a
50~\cm\ gap,\footnote{The rather large gap value reported for the hole-doped compound \cite{Li08SC} waits for an independent confirmation.} while there is some discrepancy on the existence
of further gaps. On the other hand, both investigations
carried out on thin films indicate a gap of $2\Delta/hc\approx
30$~\cm. In trying to explain this disaccord, disorder or strain
effect have to be considered. Furthermore, it is worth to note that, in all cases the further gap(s) favor to open in
individual conducting channels where the scattering rate is rather big. Thus, a universal decomposition of free carrier contribution corresponding to the Fermi-surface construction is aspired to get deeper insight of the gap analysis.

\subsection{FeTe$_{0.55}$Se$_{0.45}$ compound}

\begin{figure}[h]
 \includegraphics[width=0.8\columnwidth]{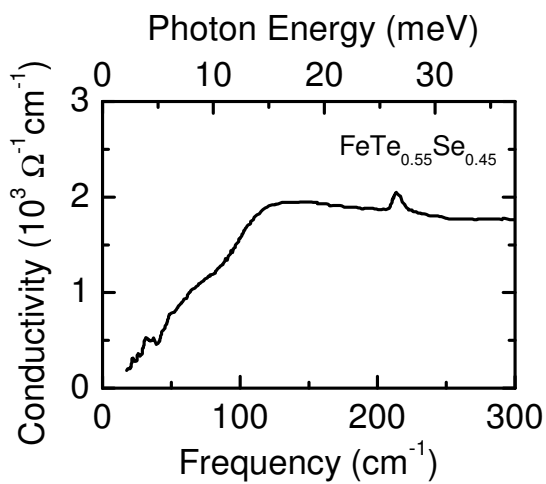}
 \caption{Optical conductivity of a FeTe$_{0.55}$Se$_{0.45}$ single crystal  ($T_c=14$~K) obtained from reflectivity measurements at $T=6$~K by  Homes {\it et al.} \cite{Homes10}. \label{fig:FeTeSe}
} \end{figure}

The so-called 11 compounds are the simplest in the iron pnictide
family, and the As-free FeSe reaches $T_c=8$~K which can be
increased to 27~K upon application of pressure. By introducing Te,
the critical temperature reaches a maximum of $T_c=14$~K for in
FeTe$_{0.55}$Se$_{0.45}$. Homes {\it et al.} have explored the
electrodynamic properties by performing reflectivity
measurements down to 6~K \cite{Homes10}. Very similar to the 122
compounds, an almost constant conductivity background is observed
in the range between 500 and 1000~\cm\ on top of which a
Drude-like component develops as the temperature is reduced. The
authors fit their data by the Drude-Lorentz approach
\cite{DresselGruner02} with a strong mid-infrared excitation which
contains a considerable tail all the way down to the lowest
frequencies. In the superconducting state at $T=6$~K  the optical
conductivity is reduced below 120~\cm\ with a prominent shoulder
at 60~\cm\ as plotted in Fig.~\ref{fig:FeTeSe}. Below the maximum
in $\sigma(\omega)$ the conductivity drops in a linear way towards
zero. It is interesting to note that only a quarter of the free
charge carriers collapse into the condensate leading to
$\lambda=5300$~\AA. Although a Drude fit of the normal state
yields $1/(2\pi c \tau) \approx 30$~\cm, the material is considered in the
dirty limit as the frequency dependent scattering rate
$1/\tau(\omega)$ is enlarged in the region of the gaps.

The data can be fitted by the Mattis-Bardeen expression
\cite{Mattis58} using a single gap of  $2\Delta_0=10.2$~meV (corresponding to 82~\cm), but
the remaining low-lying excitation call of a two-gap fit with
$2\Delta_0^{(1)}=5$~meV and $2\Delta_0^{(2)}=10.2$~meV, very similar to the
approach Wu {\it et al.} used in the electron doped 122 compounds
\cite{Wu10b}. Some in-gap excitations remain, far more than
expected from a simple $s$-wave gap at temperatures below $T_c/2$.
Nevertheless Homes {\it et al.} conclude that the energy gaps in
this material are isotropic and see no indications of nodes in the
order parameter \cite{Homes10}. The different gaps belong to
different bands crossing the Fermi surface, although a clear
assignment cannot be made from optical experiments. Since the gaps
open simultaneously with decreasing temperature below $T_c=14$~K,
the coupling constants are different $2\Delta_0/k_BT_c=4$ and 8.4.

\section{Comparison with MgB$_2$}

Is is worth to have a closer look at optical measurements on
MgB$_2$, which is a well-known two-band superconductor
\cite{Kuzmenko07}. In general a gap value corresponding to the
$\pi$ bands of approximately $2\Delta_{\pi}/hc = 25-40$~\cm\ is
extracted
\cite{Gorshunov01,Pronin01,Kaindl02,Perucchi02,Ortolani05},
corresponding to $2\Delta_{\pi}/k_BT_c\approx 1-2$. However, these
infrared reflection and transmission experiments could not see any
indication of the larger $2\Delta_{\sigma}$, expected around
110~\cm\  based on  tunneling \cite{Iavarone02}, ARPES
\cite{Souma03}, and Raman \cite{Quilty02} investigations. Only recently,
measurements on thin films with low and high impurity levels give some indications of two gaps in the optical response \cite{Ortolani08}.  Commonly
the infrared data are analyzed by  assuming that the charge
carriers in the $\pi$ and $\sigma$ bands respond independently to
the external radiation; each component is characterized by its own
plasma frequency $\omega_p$, scattering rate $1/\tau$ and
superconducting gap $2\Delta$. Although a suitable choice of
parameters allows one to fit the data, the description remain
unsatisfactory because a rise in reflectivity should always occur
at the value of the larger gap.

\section{Conclusion and Outlook}
The comparison of the various optical results on single crystals
and thin films of iron pnictides give a very similar overall
picture. The optical conductivity in the normal state consists of
some mid-infrared bands and a quasi-free carrier contribution at
$\omega=0$, which cannot be described by one simple Drude term.
Frequently the optical conductivity is decomposed  into two
independent contributions (except for the FeSe compound where only
one Drude was used), albeit no obvious assignment is possible to
the bands crossing the Fermi surface. Entering the superconducting
state the conductivity drops as the electronic density of states
becomes gapped.

No agreement exists as far as the number of gaps is concerned.
Mainly based on suggestions from other experimental methods and
theory, gaps of different size are tested. The common approach is
that simultaneously below $T_c$, each of the conductivity
contributions become completely gapped with different coupling
strength $2\Delta/k_BT_c$. Care has to be taken not to
overinterprete the conductivity spectra above 100 or 200~\cm\
since correlation and bandstructure effect as well as phonons
might cause deviations form a simple Drude behavior.

Looking in more detail at the optical conductivity  obtained by
different groups for $T<T_c$, however, significant deviations
become obvious, in particular at the low-frequency end of the
spectrum. By now, only simple $s$-wave scenarios have been
considered with no $k$-dependent order parameter. In a detailed
analysis of the optical conductivity of iron pnictides, Carbotte
and Schachinger \cite{Carbotte10} pointed out that in an extended
$s$-wave scenario gap nodes may exist in certain directions  on
the Fermi surface. These nodes can be lifted by increasing
disorder leading to a finite gap in all momentum directions.
Further low-frequency experiments have to be performed, well below
50~\cm, in order to clarify this issue.

\section*{Acknowledgements}
We appreciate discussions with D. N. Basov, S.-L. Drechsler, C.
Haule, D. van der Marel, and E. Schachinger. N.B. and D.W.
acknowledge a fellowship of the Alexander von Humboldt-Foundation.
The work in Stuttgart has been performed on samples from Zhejiang
University in Hangzhou and IFW in Dresden. We thank G. H. Cao, N.
Drichko, A. Faridian, P. Gegenwart, S. Haindl, B. Holzapfel K.
Iida, H.S. Jeevan,  P. Kallina, F. Kurth, L. J. Li, X. Lin, L.
Schultz, A. A. Voronkov, N. L. Wang, and Z. A. Xu for the
collaboration.

\end{document}